# Experimental Characterization of Three-Band Braid Relations in Non-Hermitian Acoustic Systems

Qicheng Zhang, Luekai Zhao, Xun Liu, Xiling Feng, Liwei Xiong, Wenquan Wu, and Chunyin Qiu*

Key Laboratory of Artificial Micro- and Nano-Structures of Ministry of Education and School of Physics and Technology

Wuhan University, Wuhan 430072, China

* To whom correspondence should be addressed: cyqiu@whu.edu.cn

*Abstract*. The nature of complex eigenenergy enables unique band topology to the non-Hermitian (NH) lattice systems. Recently, there has been a fast growing interest in the elusive winding and braiding topologies of the NH single and double bands, respectively. Here, we explore the even more intricate NH multi-band topology and present the first experimental characterization of the three-band braid relations by acoustic systems. Based on a concise tight-binding model, we design a ternary cavity-tube structure equipped with a highly controllable unidirectional coupler, through which the acoustic NH Bloch bands are experimentally reproduced in a synthetic dimension. We identify the NH three-band braid relations from both the perspectives of eigenvalues and eigenstates, including a noncommutative braid relation $\sigma_1 \sigma_2 \neq \sigma_2 \sigma_1$ and a swappable braid relation $\sigma_1 \sigma_2 \sigma_1 = \sigma_2 \sigma_1 \sigma_2$. Our results could promote the understanding of NH Bloch band topology and pave the way toward designing new devices for manipulating acoustic states.

*Introduction.* Topological band theory, as one of the cornerstones in condensed matter physics, provides a unified framework for classifying distinct topological phases of matter [1,2]. Recently, the band topology in non-Hermitian (NH) lattices has attracted fast growing attention since it significantly broadens and deepens our understanding to the conventional topological band theory [3-20]. Unlike the Hermitian band topology defined only by Bloch wave functions, the NH one can be defined either by Bloch wave functions or by non-Bloch wave functions, because of the strikingly different energy spectra under periodic and open boundary conditions. This yields Bloch or non-Bloch NH band theory. The former, still employing the real-valued wave vector and conventional Brillouin zone (BZ), classifies the nontrivial topology of complex band structures [13-20], while the latter enables to interpret unique open-boundary phenomena (e.g. skin effects and NH bulk-boundary correspondence) through introducing complex-valued wave vectors and a generalized BZ [3-12].

In the context of NH Bloch band, even a single band can be topologically nontrivial by generating complex-energy windings [21-24] and vortices [25,26]. While considering two bands, the systems can further exhibit bands braiding and states swapping [2,27-29], both of which, as the manifestations of NH Hamiltonians living on Riemann manifolds, directly connect with the topology of encircling exceptional points (EPs) [2,30-36]. As for NH multi-band systems, even more distinctive topological properties arise since (i) they permit the existence of multiple and higher-order EPs [37-41]; (ii) their homotopy group $\mathbb{B}_N$ ($N > 2$) is non-Abelian and enjoys novel braid relations [42-44]. Very recently, experimental realizations have been reported for the single-band winding topology [23,24] and two-band braiding topology [27-29]. However, there is no direct experimental progress on the NH multi-band topology [45-47].

In this Letter, we report the first experimental characterization of the NH three-band braid relations by acoustic systems. Theoretically, we propose a NH three-band lattice model, in which a series of band braids and state permutations are revealed from the topology of encircling multiple EPs. Experimentally, we use the synthetic dimension technique [23,27] to realize the model through a ternary-cavity structure equipped with a well-controlled unidirectional coupler (UC). Typically, a noncommutative braid relation (NBR) and a swappable braid relation (SBR) are unambiguously captured not only by retrieving the complex band structures from the effective Hamiltonian, but also by directly observing the permutations of the acoustic states. Our findings demonstrate the NH multi-band topology in a fundamental manner that has no analog to the Hermitian one, and enable novel properties and functionalities for a wide range of acoustic applications.

*Three-band braid relations and lattice model.* Here, we introduce the NH three-band braid relations governed by the braid group $\mathbb{B}_3$ [42]. As depicted in Fig. 1, each band braid in the complex *E-k* space is specified by a braid word, a product of the braid elements $\sigma_1$ and $\sigma_2$. Specifically, $\sigma_1$ ($\sigma_2$) defines an anticlockwise braid where the first (second) band crosses over the second (third) band [Fig. 1(a)], resulting in an exchange of the eigenenergies $E_1$ ($E_2$) and $E_2$ ($E_3$). Note that the eigenenergies are sorted by their real parts. Intriguing braid relations arise when we alternatively build up the band braids with $\sigma_1$ and $\sigma_2$, as shown in Figs. 1(b) and 1(c). In terms of permuting eigenenergies, the braids $\sigma_1 \sigma_2$ and $\sigma_2 \sigma_1$ exhibit an inequivalent braiding consequence, i.e., $\sigma_1 \sigma_2 \neq \sigma_2 \sigma_1$ (dubbed NBR). More specifically, $\sigma_1 \sigma_2$ gives rise to $[E_1, E_2, E_3] \to [E_2, E_3, E_1]$, whereas $\sigma_2 \sigma_1$ results in $[E_1, E_2, E_3] \to [E_3, E_1, E_2]$. In contrast, the braids $\sigma_1 \sigma_2 \sigma_1$ and $\sigma_2 \sigma_1 \sigma_2$ present an equivalent braiding consequence despite the two elements are swapped, i.e., $\sigma_1 \sigma_2 \sigma_1 =$ $\sigma_2\sigma_1\sigma_2$. We refer to it as the SBR. In this case, both $\sigma_1\sigma_2\sigma_1$ and $\sigma_2\sigma_1\sigma_2$ result in $[E_1, E_2, E_3] \to [E_3, E_2, E_1]$.

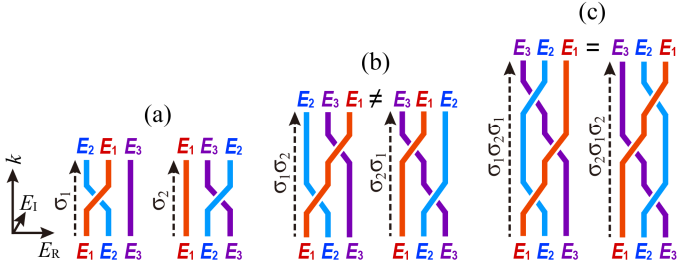

FIG. 1. Fundamental braid relations of NH three-band systems. The red, blue and purple lines represent three separable complex energy bands braided in $(E_R, E_I, k)$ space. (a) Two braid elements $\sigma_1$ and $\sigma_2$. (b) Noncommutative braid relation (NBR) $\sigma_1\sigma_2 \neq \sigma_2\sigma_1$, where $\sigma_1\sigma_2$ and $\sigma_2\sigma_1$ give rise to inequivalent braiding consequences in terms of eigenenergy permutations. (c) Swappable braid relation (SBR) $\sigma_1\sigma_2\sigma_1 = \sigma_2\sigma_1\sigma_2$, where $\sigma_1\sigma_2\sigma_1$ and $\sigma_2\sigma_1\sigma_2$ induce equivalent braiding consequences.

All the above band braids can be realized simultaneously in a single three-band lattice model. As shown in Fig. 2(a), the orbitals 1, 2 and 3 (of zero onsite energy) are coupled by a reciprocal intracell hopping $t_0$, and two nonreciprocal intercell hoppings $t_m$ and $t_n$ that span $m$ and $n$ lattices, respectively. For simplicity, we assume $t_0 = 1$, $t_m, t_n \in \mathbb{R}$, and $m, n > 0$. The Hamiltonian of this lattice model reads

$$\mathbf{H}(k) = \begin{pmatrix} 0 & t_0 & 0 \\ t_0 & 0 & t_0 \\ t_m e^{imk} + t_n e^{ink} & t_0 & 0 \end{pmatrix}, \quad (1)$$

from which one may achieve various NH three-band braiding structures. To reflect the overall braiding degree of the three separable NH bands, we introduce an integer topological invariant $v$ [27],

$$v := \sum_{i,j=1}^{3} \frac{1}{2\pi i} \oint_{BZ} \frac{d\ln\tilde{E}_{ij}}{dk} dk, \quad (2)$$

where $\tilde{E}_{ij}(k) = [E_i(k) - E_j(k)]/2$ with $E_i$ and $E_j$ ($i \neq j$) are two eigenvalues of $\mathbf{H}(k)$. Remarkably, the nonreciprocal long-range couplings designed in our model ensure that all the braids are composed of the anticlockwise elements $\sigma_1$ and $\sigma_2$, and hence the braiding degree $v$ equals to the total number of $\sigma_1$ and $\sigma_2$ (see Supplemental Material [48]). This guides us to seek the desired braids from the phase diagram of $v$. Here, we consider the particular case of $\mathbf{H}(k)$ under $m = 1$ and $n = 2$, whose phase diagram is shown in Fig. 2(b). All the band braids sketched in Fig. 1 are achieved, with the dots corresponding to the parameters to be realized in our follow-up experiments. Note that the braid words, which are defined in an intact BZ $[k_0, k_0 + 2\pi]$, depend on the selection of the initial momentum $k_0$ ($-\pi/6$ throughout the work) [20]. An arbitrary selection of $k_0$ may influence the order of $\sigma_1$ and $\sigma_2$ for a given set of parameters.

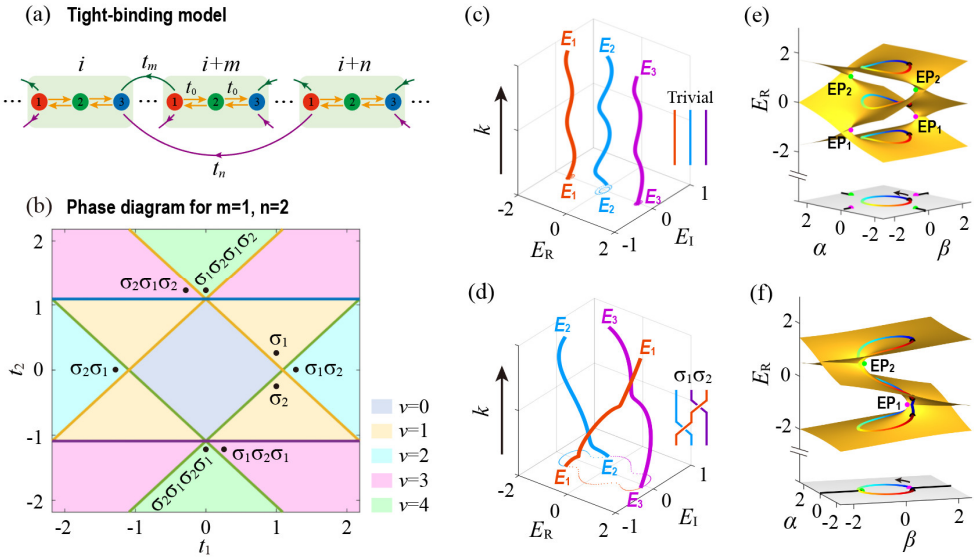

FIG. 2. 1D NH lattice model and three-band braids. (a) A sketch of the three-band lattice model. In addition to the intracell hopping $t_0$, it features two nonreciprocal intercell hoppings $t_m$ and $t_n$ that span $m$ and $n$ lattices, respectively. (b) Phase diagram for $m = 1$ and $n = 2$. The dots mark the band braids to be realized in our experiments. (c) and (d): A trivial and a nontrivial three-band braids, respectively. (e) and (f): The associated manifestations on Riemann surfaces (real parts), where EP$_1$ (EP$_2$) represents the EP degenerated by the two lower (higher) Riemann sheets. The eigenvalues, the EPs and their branch cuts are projected to the $\alpha$-$\beta$ plane, on which the black dot is the starting point and the arrow indicates the evolution of the momentum.

As examples, figures 2(c) and 2(d) display two representative band braids, one for trivial (with $t_1 = 0.1$ and $t_2 = 0.3$) and the other for non-trivial (with $t_1 = 1.2$ and $t_2 = 0$), respectively. In the former case, the three bands do not braid with each other since the three complex eigenvalues $E_1$, $E_2$ and $E_3$ return to themselves trivially over the shifted BZ. In the latter case, however, the eigenvalues permute and none of them come back to themselves, i.e., $[E_1,E_2,E_3] \to [E_2,E_3,E_1]$, a manifestation of the Bloch band braid $\sigma_1\sigma_2$. The nontrivial eigenvalue permutation can be traced back to the underlying topology of encircling EPs on Riemann surfaces. To demonstrate that, we expand the $k$-space Hamiltonian $\mathbf{H}(k)$ into a parametric-space Hamiltonian $\mathbf{H}(\alpha,\beta)$ by the substitutions $\cos k \to \alpha$ and $\sin k \to \beta$, so that the shifted BZ is mapped to a unit circle $\alpha^2 + \beta^2 = 1$ in $(\alpha,\beta)$-space. As shown in Figs. 2(e) and 2(f), the eigenvalues of $\mathbf{H}(\alpha,\beta)$ define the three sheets of Riemann surfaces, and the eigenvalues of $\mathbf{H}(k)$ map out the colored loops on the sheets. When the loops do not encircle any EPs, the three eigenvalues of $\mathbf{H}(k)$ do not intersect and never exchange [Fig. 2(e)]. By contrast, encircling the EPs inevitably crosses the branch cuts of the Riemann surfaces, leading to the permutation of eigenvalues. For instance, encircling EP$_1$ and EP$_2$ in sequence [Fig. 2(f)] results in the permutation process $[E_1,E_2,E_3] \to [E_2,E_1,E_3] \to [E_2,E_3,E_1]$, yielding the band braid $\sigma_1\sigma_2$ [Fig. 2(d)]. More eigenvalue permutations and associated topologies of encircling EPs are provided in *Supplemental Material* [48]. Note that our three-band model with versatile EP-topology is markedly different from the previous two-band models [2,20,27-29], in which the two-eigenvalue permutation can always be treated as a trivial process in terms of non-Abelian effects.

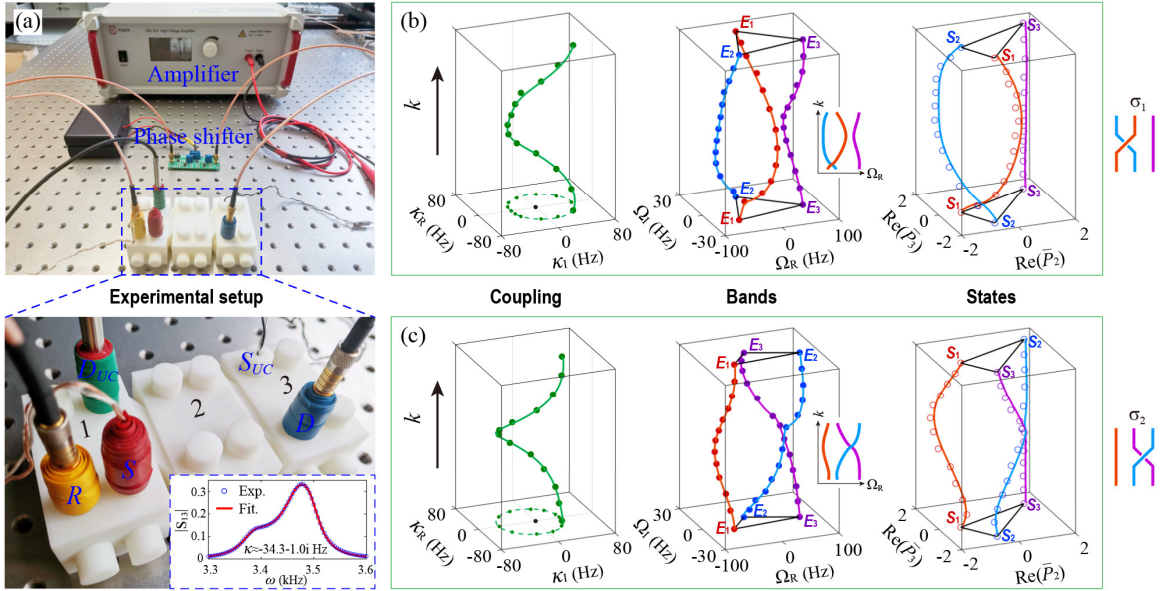

FIG. 3. Experimental setup and acoustic realization of the two elementary braids $\sigma_1$ and $\sigma_2$. (a) Experimental setup. The acoustic cavities 1, 2 and 3 emulate three orbitals and the narrow tubes in between mimic the reciprocal coupling $t_0$. A controllable unidirectional coupler (UC), consisting of a microphone $D_{UC}$, an amplifier, a phase shifter, and a loudspeaker $S_{UC}$, is introduced between the cavities 1 and 3 to generate the unidirectional complex coupling $\kappa = \rho e^{i\theta}$. A source $S$ and a detector $D$ are used to excite and detect acoustic transmission responses, respectively, together with a detector $R$ applied for phase reference. The inset exemplifies how the complex coupling $\kappa$ is retrieved by fitting the transmission response $|S_{13}(\omega)|$. (b)-(c) Experimental results for the elementary braids $\sigma_1$ and $\sigma_2$, including the designed $\kappa$ (left), the band braids (middle) and their projections on the $\Omega_R$-$k$ plane (insets), and the state permutations (right). To facilitate observation, the initial and final eigenfrequencies (states) are linked into triangles. All experimental data (dots/circles) match well the theoretical predictions (lines).

*Experimental characterization of the three-band braid relations.* Notice that the braiding topology of NH Bloch bands cannot be identified in a finite lattice with open boundaries, whose eigenspectra are qualitatively different from those of the corresponding infinite system [2,3,23,27,29]. Here we use the concept of *synthetic dimension* to implement the 1D NH model, by designing an acoustic cavity-tube structure equipped with a unidirectional coupler (UC). As displayed in Fig. 3(a), the system consists of three identical air cavities with a complex dipole resonance frequency $\Omega_0 \approx 3450 - 24.4i$ Hz. The narrow tubes connecting two adjacent cavities produce a reciprocal intracell coupling $t_0 \approx 43.3$ Hz. These intrinsic parameters are retrieved by fitting the transmission response $|S_{13}(\omega)|$ of the system in the absence of UC, where the subscripts $i$ and $j$ in $S_{ij}(\omega)$ denote the cavities inserted with the acoustic detector $D$ and source $S$, respectively. On the other hand, a UC that consists of a microphone $D_{UC}$, an amplifier, a phase shifter, and a loudspeaker $S_{UC}$, is introduced between the cavities 1 and 3 for achieving the unidirectional coupling $\kappa = \rho e^{i\theta}$. Notice that the amplitude $\rho$ and phase $\theta$ can be controlled by the amplifier and phase shifter, respectively [29]. Thus we can use $\kappa$ to realize the term $t_m e^{imk} + t_n e^{ink}$ of $\mathbf{H}(k)$ in the $k$-resolved synthetic dimension. Turning on the UC and using the already assessed intrinsic parameters, the value of $\kappa$ can be precisely retrieved by fitting the transmission response $|S_{13}(\omega)|$, as exemplified in the inset of Fig. 3(a). Eventually, we obtain the complex acoustic band structures by substituting all the fitted parameters into $\mathbf{H}(k)$ and solving the $k$-resolved eigenfrequencies $\Omega = \Omega_R + i\Omega_I$. Meanwhile, we extract the real-part information of the eigenstates, $\text{Re}\psi_i = [1, \text{Re}\bar{P}_2, \text{Re}\bar{P}_3]^T$, by measuring the transmission responses $P_1 = S_{13}(\Omega_R)$, $P_2 = S_{23}(\Omega_R)$ and $P_3 = S_{33}(\Omega_R)$ at real eigenfrequency $\Omega_R$, where $\bar{P}_2 = P_2/P_1$ and $\bar{P}_3 = P_3/P_1$. More experimental and sample details are presented in Supplemental Material [48]. Next, we manifest the topological band braids and their relations, from both the synthesized band structures and the acoustic state permutations.

Figures 3(b) and 3(c) show our experimental results for the two elementary band braids $\sigma_1$ and $\sigma_2$. Firstly, we tune the UC to ensure the unidirectional coupling $\kappa(k) \approx 1.0 t_0 e^{ik} + 0.2 t_0 e^{i2k}$, at a momentum step of $\pi/6$ [Fig. 3(b), left]. In the 3D space spanned by $(\Omega_R, \Omega_I, k)$, the red and blue bands twist around each other, and the purple band stays out of the twisting [Fig. 3(b), middle]. As a visual manifestation of the band braid $\sigma_1$, the eigenfrequencies $E_1$ and $E_2$ permute as $k$ evolves over a shifted BZ. The permutation can be further confirmed by the acoustic states evolved in $[\text{Re}\bar{P}_2, \text{Re}\bar{P}_3, k]$ space, where the states $S_1$ and $S_2$ swap their positions [Fig. 3(b), right]. In a similar manner, the acoustic band braid $\sigma_2$ is realized by tuning $\kappa(k) \approx 1.0 t_0 e^{ik} - 0.2 t_0 e^{i2k}$. As shown in Fig. 3(c), the permutation between $E_2(S_2)$ and $E_3(S_3)$ can be clearly observed. Note that the complex eigenfrequency $\Omega$ in Fig. 3 has been deducted by $\Omega_0$ for clarity. The concrete values of eigenfrequencies and states are provided in Supplemental Material [48].

bands get entangled and all three eigenfrequencies permute with $k$ evolving. Their braid words can be respectively recognized as $\sigma_1\sigma_2$ [Fig. 4(a)] and $\sigma_2\sigma_1$ [Fig. 4(b)] via the projected band structures (insets). Under the braid $\sigma_1\sigma_2$, the three eigenfrequencies change from $[E_1, E_2, E_3]$ into $[E_2, E_3, E_1]$, and the corresponding states permute from $[S_1, S_2, S_3]$ into $[S_2, S_3, S_1]$. The braid $\sigma_2\sigma_1$, however, undergoes an entirely different permutation process, i.e., $[E_1, E_2, E_3] \to [E_3, E_1, E_2]$ and $[S_1, S_2, S_3] \to [S_3, S_1, S_2]$. Both the permutations of the eigenvalues and eigenstates directly witness the inequivalent braiding consequences of $\sigma_1\sigma_2$ and $\sigma_2\sigma_1$, exhibiting a character of non-Abelian effect. Different from the non-Abelian characteristics unveiled for three NH states evolving in real space [49-51] or parametric space [45,46], here we aim to the braiding topology in the momentum space, which is of great significance for understanding the fundamental Bloch band theory in NH lattice systems. Note that both the braids $\sigma_1\sigma_2$ and $\sigma_2\sigma_1$ have an even permutation parity associated to the global biorthogonal Berry phase $Q = 0$ [20]. However, their inequivalent braiding consequences are still characterized by the unequal non-Abelian Berry phases [52], which can be explicitly described by the unitary matrices [0 1 0; 0 0 1; 1 0 0] and [0 0 1; 1 0 0; 0 1 0], respectively.

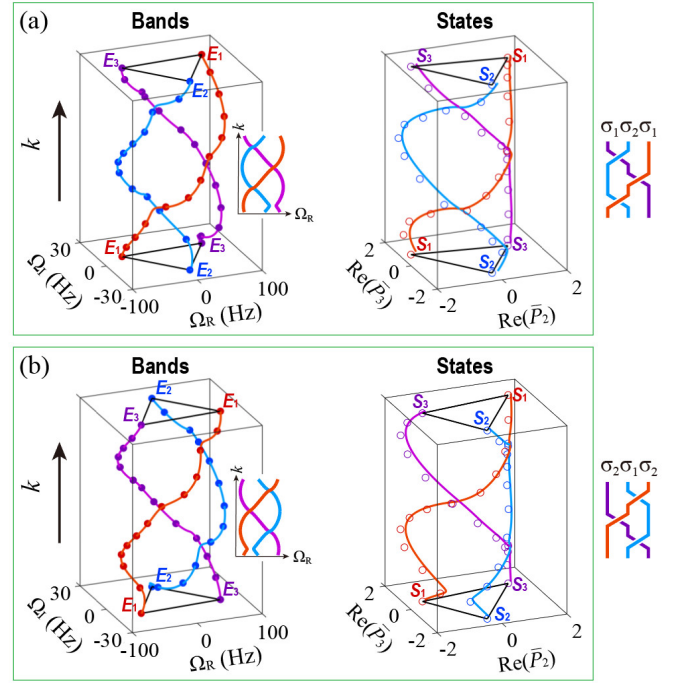

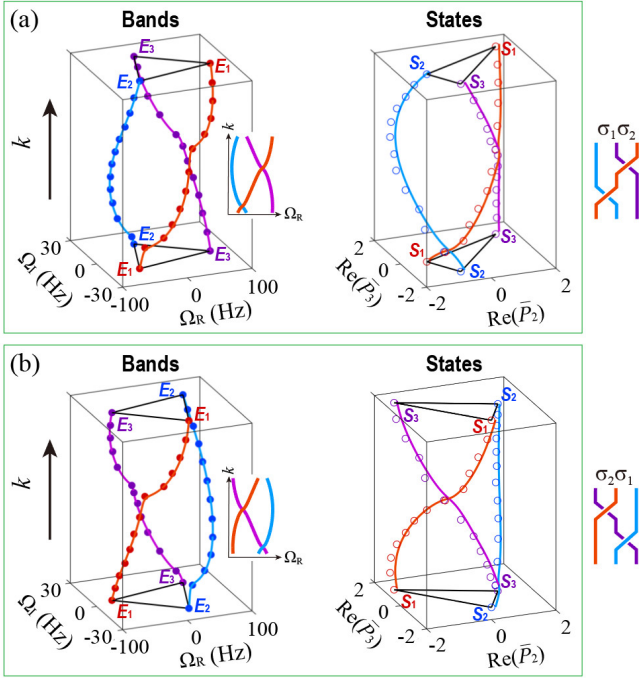

FIG. 4. Experimental characterization of the NBR $\sigma_1\sigma_2 \neq \sigma_2\sigma_1$. (a)-(b) Results of the band braids $\sigma_1\sigma_2$ and $\sigma_2\sigma_1$. Left, the acoustic complex band structures and their projections (insets); right, the acoustic state permutations. The braids $\sigma_1\sigma_2$ and $\sigma_2\sigma_1$ generate inequivalent braiding consequences.

Now we experimentally characterize the NBR through controlling $\kappa(k)$ to map the points $(t_1, t_2) = (1.2, 0)$ and $(t_1, t_2) = (-1.2, 0)$ in the phase diagram (see details in Supplemental Material [48]). In this case, all three acoustic

FIG. 5 Experimental characterization of the SBR $\sigma_1\sigma_2\sigma_1 = \sigma_2\sigma_1\sigma_2$. (a)-(b) Results of the band braids $\sigma_1\sigma_2\sigma_1$ and $\sigma_2\sigma_1\sigma_2$. Left, the acoustic complex band structures and their projections (insets); right, the acoustic state permutations. The braids $\sigma_1\sigma_2\sigma_1$ and $\sigma_2\sigma_1\sigma_2$ result in equivalent braiding consequences.

Figure 5 shows our experimental results for characterizing the SBR $\sigma_1\sigma_2\sigma_1 = \sigma_2\sigma_1\sigma_2$. To reproduce such band braids, we tune the $\kappa(k)$ to map the points $(t_1, t_2) = (0.3, -1.2)$ and $(t_1, t_2) = (-0.3, 1.2)$ in the phase diagram (see Supplemental Material [48]). As predicted, the three acoustic bands twist in a more intricate way, forming two band braids whose elements are swapped i.e., $\sigma_1\sigma_2\sigma_1$ [Fig. 5(a)] and $\sigma_2\sigma_1\sigma_2$ [Fig. 5(b)], respectively. In a stark contrast to the NBR, the braids $\sigma_1\sigma_2\sigma_1$ and $\sigma_2\sigma_1\sigma_2$ have equivalent braiding consequences in terms of permuting eigenvalues and eigenstates: $[E_1, E_2, E_3] \to [E_3, E_2, E_1]$ and $[S_1, S_2, S_3] \to [S_3, S_2, S_1]$. In this case, the two braids have the same global biorthogonal Berry phase $Q = \pi$ and non-Abelian Berry phase [0 0 1; 0 1 0; 1 0 0]. The former indicates the same odd permutation parity, while the latter evidences the equivalence in braiding consequence.

*Conclusions and discussions*. From both the perspectives of eigenfrequencies and states, we have experimentally characterized two fundamental braid relations for the multi-band braiding topology. We have not only demonstrated a variety of acoustic band braids in synthetic dimension (e.g., $\sigma_1$, $\sigma_2$, $\sigma_1\sigma_2$, $\sigma_2\sigma_1$, $\sigma_1\sigma_2\sigma_1$ and $\sigma_2\sigma_1\sigma_2$), but also achieved all possible permutation consequences for the three-state braiding systems. The latter means that an input acoustic state can be manipulated into any of the six output states by designing the band braids, which could advance the applications on acoustic logic gates and switches.

Interestingly, based on the two fundamental braid relations, we can further deduce a generic braid relation for two longer but distinct braid sequences consisting of the elements $\sigma_1$ and $\sigma_2$ alternatively: if the total number of the braiding elements is a multiple of three, their braiding consequences are equivalent; otherwise inequivalent (see Supplemental Material [48] for the proof). The latter is experimentally evidenced by the two braids $\sigma_1\sigma_2\sigma_1\sigma_2$ and $\sigma_2\sigma_1\sigma_2\sigma_1$ allowed in our current model [48]. Moreover, our work can be extended to demonstrate the collective behaviors (e.g., creation, split, braiding, annihilation, coalescence, etc.) of multiple (or higher-order) EPs and associated band topologies [41,44].

## ACKNOWLEDGEMENTS

This work is supported by the National Natural Science Foundation of China (Grant No. 11890701, 12104346, 12004287), and the Young Top-Notch Talent for Ten Thousand Talent Program (2019-2022).